\documentstyle[12pt]{article}
\begin{document}
\begin{titlepage}

\renewcommand\baselinestretch {1,5}
\large\normalsize

\begin{center}
National Academy of Sciences of Ukraine\\
Institute for Condensed Matter Physics
\end{center}

\vspace{3cm}
\hspace{8cm}
\begin{tabular}{l}
Preprint\\
ICMP-94-15E
\end{tabular}

\vspace{2cm}

\begin{center}
{\large {O.Derzhko, T.Krokhmalskii\\}}
\vspace{1cm}
{\Large {Random $s=1/2$
$XY$ chains and the theory of quasi-one-dimensional ferroelectrics
with hydrogen bonds}}
\end{center}

\vspace{6cm}

\begin{center}
L'viv-1994\\
\end{center}

\end{titlepage}

\clearpage
\renewcommand\baselinestretch {1,5}
\large\normalsize



\noindent
O.Derzhko, T.Krokhmalskii\\
Random $s=1/2$
$XY$ chains and the theory of quasi-one-dimensional ferroelectrics
with hydrogen bonds\\
\\
The paper presents a new numerical approach for studying
the thermodynamical and dynamical properties of finite
spin-$\frac{1}{2}$ $XY$ chains.
Special attention is given to examining
the influence of
disorder on the average transverse dynamical susceptibility
of Ising chain in random transverse field.

\vspace{2cm}

\noindent
\copyright 1994

\clearpage

\renewcommand\baselinestretch {1,5}
\large\normalsize

The present paper deals with the equilibrium properties of
spin-$\frac{1}{2}$ non-uniform anisotropic $XY$ chain
in transverse field that is
defined by the Hamiltonian
\begin{eqnarray}
H=\sum_{j=1}^{N}\Omega_js_j^z+\sum_{j=1}^{N-1}(J^{xx}_js^x_js^x_{j+1}+
J^{yy}_js^y_js^y_{j+1}).
\end{eqnarray}
The quantities of main interest are the time-dependent correlation
functions
$<s^{\alpha}_j(t)s^{\beta}_{j+n}>$,
$<(...)> \equiv {\mbox {Sp}}[{\mbox {e}}^{-\beta H}(...)]
/{\mbox {Sp}}{\mbox {e}}^{-\beta H}$,
$s^{\alpha}_j(t) \equiv {\mbox {e}}^{iHt}s^{\alpha}_j
{\mbox {e}}^{-iHt}$
and dynamical susceptibilities
$\chi_{\alpha \beta}(\ae, \omega) \equiv
\sum_{n}{\mbox {e}}^{i \ae n}
\int_0^{\infty} dt {\mbox {e}}^{i( \omega +i \epsilon ) t}
\frac{1}{i} <[s^{\alpha}_j(t),$
$s^{\beta}_{j+n}]>$.

The system considered in partial cases has well-known ferroelectric
interpretation in connection with some quasi-one-dimensional
hydrogen-bonded ferroelectric crystals like
$Cs(H_{1-x}D_x)_2PO_4$,
$PbH_{1-x}D_xPO_4$ etc.
Indeed, in such compounds
the hydrogen bonds, that play an important role for the
ferroelectricity, run along chains, the intrachain interactions are
much stronger than the interchain ones, and a suitable model for
describing protons behaviour in a chain is
spin-$\frac{1}{2}$
Ising model in transverse
field, which is contained in (1) ($J^{yy}_j=0$, $J^{xx}_j$
is interaction between neighbouring hydrogen bonds along chain
and $\Omega_j$
describes proton tunneling motion).
The non-uniformity may be caused physically by deuterization of some
hydrogen bonds. Under simplest assumption one believes
that the substitution
$H \rightarrow D$ changes significantly the tunneling probability and
practically does not affect the intersite interaction parameters.
For this case one should put
$J^{xx}_j=J$ and suppose that
$\Omega_j$ varies randomly from
site to site
taking  value $\Omega $ with probability $1-x$ and
zero value with probability $x$,
that is, one should consider random chain (1) with
the probability distribution density
\begin{eqnarray}
p(...,\Omega_j, J_j^{xx}, J_j^{yy},...)=
\prod_{j=1}^N[x\delta (\Omega_j)+
(1-x)\delta (\Omega_j-\Omega)]\delta (J_j^{xx}-J)\delta (J_j^{yy}).
\end{eqnarray}
Statistical mechanics calculations for this model bear strong relevance
for
paraelectric phase properties of mentioned crystals and have been
performed
mainly in the limiting cases
$x=1$ and $x=0$
(see e.g.$^{1-3}$
and references therein).
Such studies were encouraged by the fact that the model (1) with the
help of Jordan-Wigner transformation can be rewritten as a system of
non-interacting fermions and thus a lot of result can be obtained
exactly$^4$.
Nevertheless, there are serious difficulties in calculation
of longitudinal time-dependent correlation function $<s^x_j(t)s^x_{j+n}>$
and therefore the dynamical susceptibility
$\chi_{xx}(\ae, \omega)$,
and the complete solution even for perfect (non-random) case
has never been found. The results obtained within different approximate
approaches have serious shortcomings.

In our previous
papers$^{5-7}$
we suggested a new numerical approach for evaluation of
thermodynamical properties and both equal-time and different-time spin
correlation functions
for the model
defined by (1).
In the present paper the proposed approach is described in more detail
and is applied for
examining the influence of disorder on the average transverse
dynamical susceptibility of
spin-$\frac{1}{2}$ Ising chain in random transverse field (1), (2).
Such results, up to the best authors' knowledge, are obtained for the
first time.

In order to obtain the equilibrium statistical properties of the system in
question (1) one should first perform Jordan-Wigner transformation from
operators $s^\pm _j \equiv s^x_j\pm is^y_j$ to Fermi operators
$c_1=s^-_1, c_j=
\prod_{n=1}^{j-1}(-2s_n^z)s^-_j,\; j=2,...,N$
and then to diagonalize the obtained quadratic form by a canonical
transformation $\eta_k=\sum_{j=1}^N(g_{kj}c_j+h_{kj}c_j^+)$ with the outcome
$H=\sum_{k=1}^N\Lambda_k(\eta^+_k\eta_k-\frac{1}{2})$. The elementary
excitation
spectrum $\Lambda_k$ and the coefficients $g_{kn} \equiv
(\Phi_{kn}+\Psi_{kn})/2,\; h_{kn} \equiv (\Phi_{kn}-\Psi_{kn})/2$ are
determined from
\begin{eqnarray}
\Lambda_k\Phi_{kn}=\sum_{j=1}^{N}\Psi_{kj}({\bf A}+{\bf B})_{jn},
\nonumber\\
\Lambda_k\Psi_{kn}=\sum_{j=1}^{N}\Phi_{kj}({\bf A}-{\bf B})_{jn},
\end{eqnarray}
where
$A_{ij} \equiv \Omega_i \delta_{ij}+ \frac{J^{xx}_i+J^{yy}_i}{4}
\delta_{j,i+1} + \frac{J^{xx}_{i-1}+J^{yy}_{i-1}}{4} \delta_{j,i-1}$,
$B_{ij} \equiv \frac{J^{xx}_i-J^{yy}_i}{4} \delta_{j,i+1} -
\frac{J^{xx}_{i-1}-J^{yy}_{i-1}}{4} \delta_{j,i-1}$.
Eqs. (3) reduce to standard problems
\begin{eqnarray}
\Lambda_k^2\Phi_{kn}=
\sum_{j=1}^{N}\Phi_{kj}[({\bf A}-{\bf B})({\bf A}+{\bf B})]_{jn},
\nonumber\\
\Lambda_k^2\Psi_{kn}=
\sum_{j=1}^{N}\Psi_{kj}[({\bf A}+{\bf B})({\bf A}-{\bf B})]_{jn}
\end{eqnarray}
for $N\times N$ five diagonal banded
matrices $({\bf A}\mp{\bf B})({\bf A}\pm{\bf B})$.

Next it is necessary to express the quantities of interest
through  $\Lambda_k$, $\Phi_{kj}$, $\Psi_{kj}$. Since the
partition function $Z(\beta , N)=
\prod_{k=1}^{N}2\cosh\frac{\beta\Lambda
_k}{2}$ involves all $\Lambda_k^2$,
the knowledge of eigenvalues of
matrices $({\bf A}\mp{\bf B})({\bf A}\pm{\bf B})$
permits one to obtain the thermodynamical properties of the model in
question (1).
Reformulating spin operators in terms of operators
$\varphi_j^+ \equiv c^+_j + c_j
= \sum_{m=1}^N \Phi_{mj} (\eta^+_m + \eta_m)$
and
$\varphi_j^- \equiv c^+_j - c_j
= \sum_{m=1}^N \Psi_{mj} (\eta^+_m - \eta_m)$,
that is, $s_j^x= \prod_{n=1}^{j-1}(\varphi_n^+ \varphi^-_n)
\varphi^+_j /2$,
$s_j^y= \prod_{n=1}^{j-1}(\varphi_n^+ \varphi^-_n) \varphi^-_j /(2i)$,
$s_j^z=- \varphi_j^+ \varphi^-_j /2$,
noting that the calculation of spin correlation functions reduces to
exploiting Wick-Bloch-de Dominicis theorem,
and calculating the elementary contractions
\begin{eqnarray}
<\varphi^+_j(t) \varphi^+_m>=\sum_{p=1}^N \Phi_{pj} \Phi_{pm}
\frac{\cosh (i\Lambda_p t - \frac{\beta \Lambda_p}{2})}
{\cosh \frac{\beta \Lambda_p}{2}},
\nonumber\\
<\varphi^+_j(t) \varphi^-_m>=\sum_{p=1}^N \Phi_{pj} \Psi_{pm}
\frac{\sinh (-i\Lambda_p t + \frac{\beta \Lambda_p}{2})}
{\cosh \frac{\beta \Lambda_p}{2}},
\nonumber\\
<\varphi^-_j(t) \varphi^+_m>=-\sum_{p=1}^N \Psi_{pj} \Phi_{pm}
\frac{\sinh (-i\Lambda_p t + \frac{\beta \Lambda_p}{2})}
{\cosh \frac{\beta \Lambda_p}{2}},
\nonumber\\
<\varphi^-_j(t) \varphi^-_m>=-\sum_{p=1}^N \Psi_{pj} \Psi_{pm}
\frac{\cosh (i\Lambda_p t - \frac{\beta \Lambda_p}{2})}
{\cosh \frac{\beta \Lambda_p}{2}},
\end{eqnarray}
one finds that
$\Phi_{kp}$, $\Psi_{kg}$ for all $\Lambda_k$ are involved in final
expressions for equal-time and different-time spin correlation
functions.
Further numerical computation
may be
performed in a following way:
first, to solve the eigenvalues and eigenvectors problem for matrix
$({\bf A}+{\bf B})({\bf A}-{\bf B})$ (4) obtaining in result
$\Lambda_k^2$ and $\Psi_{kj}$, and second, having
$\Psi_{kj}$ and
$\Lambda_k=\sqrt{\Lambda_k^2}$, to find $\Phi_{kj}$ from Eq.(3);
thermodynamics and spin correlation functions are expressed via
the sought quantities.
It should be stressed
that here appears only the eigenvalues and eigenvectors problem for
$N\times N$
five diagonal banded
matrix that is the remarkable
peculiarity of the system under consideration (1).
This is a key difference of our approach in comparison with other
finite-chain calculations (reported e.g. in$^8$)
where the numerical
diagonalization of the Hamiltonian (1) in the Hilbert space of
dimension $2^N$ is performed.

The remainder of the paper deals with numerical calculation of
transverse dynamical susceptibility
$\chi_{zz}(\ae, \omega)$
for spin-$\frac{1}{2}$ Ising chain
in random transverse field (1), (2).
Acting as described above one finds that
$4<s^z_j(t)s^z_{j+n}>=< \varphi_j^+(t) \varphi^-_j(t)
\varphi_{j+n}^+ \varphi^-_{j+n}>=
< \varphi_j^+(t) \varphi^-_j(t)><\varphi_{j+n}^+ \varphi^-_{j+n}>
-< \varphi_j^+(t) \varphi^+_{j+n}><\varphi_{j}^-(t) \varphi^-_{j+n}>
+< \varphi_j^+(t)
\varphi^-_{j+n}><\varphi_{j}^-(t) \varphi^+_{j+n}>$,
$<s^z_{j+n}s^z_j(t)>=<s^z_{j+n}(-t)s^z_j>$.
Computing then numerically the elementary contractions (5) involved
in the correlation functions
$<s^z_{100}(t)s^z_{100+n}>$,
$<s^z_{100+n}(-t)s^z_{100}>$
for the
random chain of $200$ spins
with $J=-2$, $\Omega =1$ at $\beta =20$,
performing numerically the integration over $t$ and
summation over $n$,
and averaging over the realization (typically few hundred) we
have obtained the results some of which are
presented in Fig.1.
As the concentration of the sites with transverse field $\Omega $
increases
$\chi_{zz}(\ae, \omega)$
rebuilds from the generally known
Ising-like type behaviour to the behaviour inherent to
Ising model in transverse field
derived for the first time
with periodic boundary conditions imposed in$^9$.
Note, that the depicted frequency shapes lose their symmetry and
develop into smooth curves at $x=0$.
The obtained susceptibilities exhibit a lot of structure. The detailed
structure is induced by the disordered arrangement of two values of
transverse field $0$ and $\Omega =1$.
Considering at first the case of small concentrations of $\Omega $
one can find that each well-defined peak is associated with the
susceptibility of certain chain that is
determined
by values of transverse field only in the
local environment of spin at site $j=100$ (see Figs.2-4).
While $x$ decreases the number of possible local structures (and thus the
number of peaks) increases and the peaks appear almost at all
frequencies. Nevertheless,
since the difference in their magnitudes conditioned by the probability
of their appearance is large,
even at rather small $x$ it is still possible to
recognize the peaks corresponding to some simple local structures.
Completely smooth curves appear only in the limiting case $x=0$.
It should be added that the similar calculations for some other types
of disorder yield smooth frequency shapes.
Finally, one should remind that the
obtained results are exact only for finite chains and it is necessary
to be careful while making extrapolations to $N=\infty $, but,
apparently,
the picture qualitatively will remain the same.

In conclusion it should be underlined, that
the suggested method appears to be of great use in
study of dynamical properties and random versions of
spin-$\frac{1}{2}$ $XY$ chains
and
the derived results may be useful for interpretation of
observable data in dynamical
experiments on quasi-one-dimensional compounds that can be described in frames
of model (1).

The present paper was presented at Plenary Session at
Ukrainian-Polish \& East-European Workshop on Ferroelectricity and
Phase Transitions (September 18-24, 1994, Uzhgorod - V.Remety,
Ukraine)$^10$.
The authors would like to express their gratitude to participants of the
Workshop for interest in this paper and discussions.\\
\\
\noindent
REFERENCES\\
1. J.A.Plascak, A.S.T.Pires, F.C.S\'{a} Barreto,
{\em Solid State Commun.}, {\bf 44,} 787 (1982).\\
2. S.Watarai, T.Matsubara,
{\em J. Phys. Soc. Jap.}, {\bf 53,} 3648 (1984).\\
3. R.R.Levitsky, J.Grigas, I.R.Zachek, Ye.V.Mits,
W.Paprotny, {\em Ferroelectrics}, {\bf 64,} 60 (1985).\\
4. E.Lieb, T.Schultz, D.Mattis,
{\em Ann.Phys.}, {\bf 16,} 407 (1961).\\
5. O.V.Derzhko, T.Ye.Krokhmalskii, {\em Visn.L'viv.Univ.,
ser.fiz.}, 26, 47 (1993) (in Ukrainian).\\
6. O.Derzhko, T.Krokhmalskii, {\em Ferroelectrics}, {\bf 153,} 55 (1994).\\
7. O.Derzhko, T.Krokhmalskii, {\em preprint ICMP-94-9E}
(1994).\\
8. M.D'Iorio, U.Glaus, E.Stoll, {\em Solid State
Commun.}, {\bf 47,} 313 (1983).\\
9. Th.Niemeijer, {\em Physica}, {\bf 36,} 377 (1967).
10. O.Derzhko, T.Krokhmalskii, {\em Ukrainian-Polish \& East-European
Workshop on Ferroelectricity and Phase Transitions. Abstracts.
September 18-24, 1994, Uzhgorod - V.Remety, Ukraine}, 64 (1994).

\clearpage
\noindent
{\bf Figure captions}\\

\vspace{0.25cm}

\noindent
Fig.1. Frequency-dependent real and imaginary parts of
transverse susceptibility
$\chi_{zz}(\ae ,\omega )$
for spin-$\frac{1}{2}$
Ising chain in random transverse field (1), (2) at $\ae =0$
for different values of concentration $x$;
$\epsilon =0.01$.

\vspace{1cm}

\noindent
Fig.2. $\chi_{zz}(0,\omega )$
for certain spin-$\frac{1}{2}$
Ising chains. It is determined by values of transverse field in
local environment of spin at site $j=100$ (no transverse fields)
and not denoted $\Omega_j$s do not influence it;
$\epsilon =0.01$.

\vspace{1cm}

\noindent
Fig.3. The same as in Fig.2 for one transverse
field in
local environment of spin at site $j=100$;
$\epsilon =0.01$.

\vspace{1cm}

\noindent
Fig.4. The same as in Fig.2 for two transverse
fields in
local environment of spin at site $j=100$;
$\epsilon =0.01$.

\clearpage

Address:\\
Institute for Condensed Matter Physics\\
1 Svientsitskii St., L'viv-11, 290011, Ukraine\\
Tel.: (0322)761054\\
Fax: (0322)761978\\
E-mail: icmp@sigma.icmp.lviv.ua

\end{document}